\documentclass[a4paper]{spie}  
\usepackage[]{graphicx}
\usepackage{multirow}

\title{MOSE: zooming on the Meso-NH mesoscale model performances at the surface layer at ESO sites (Paranal and Armazones)} 

\author{Franck Lascaux\supit{a}, Elena Masciadri\supit{a}
\skiplinehalf
\supit{a}INAF/Osservatorio Astrofisico di Arcetri, 5 L.go E. Fermi, 50125 Florence, Italy; \\
}

\authorinfo{Send correspondence to Franck Lascaux and Elena Masciadri.\\
E-mail: 
lascaux@arcetri.astro.it - masciadri@arcetri.astro.it.}

\begin{document} 
  \maketitle 

\begin{abstract}
In the context of the MOSE project, in this contribution we present a detailed analysis of the 
Meso-NH mesoscale model performances and their dependency 
on the model and orography horizontal resolutions in proximity of the ground. 
The investigated sites are Cerro Paranal (site of the ESO Very Large Telescope - VLT) and 
Cerro Armazones (site of the ESO European Extremely Large Telescope - E-ELT), in Chile.
At both sites, data from a rich statistical sample of different nights are available - from AWS (Automated Weather Stations) and masts - 
giving access to wind speed, wind direction and temperature at different levels near the ground (from 2~m to 30 m above the ground).
In this study we discuss the use of a very high horizontal resolution ($\Delta$X=0.1 km) 
numerical configuration that overcomes some specific limitations put in 
evidence with a standard configuration with $\Delta$X=0.5 km. 
In both sites results are very promising. 
The study is co-funded by ESO and INAF.
\end{abstract}



\section{INTRODUCTION}
\label{sec:intro}  
This work is part of the MOSE (MOdeling ESO Sites) project, a feasibility study for the turbulence 
prediction at two ESO Sites: Cerro Paranal (site of the VLT) and Cerro Armazones (future site of the E-ELT).
One of the main goal of the MOSE project is to supply a tool for optical turbulence forecasts to support the scheduling 
of the scientific programs and the use of AO facilities at the VLT and at the E-ELT.
In a joint paper, Masciadri \& Lascaux \cite{mas12} have presented the results of 
numerical simulations of 20 different summer nights (in November and December 2007) with the mesoscale model Meso-NH \cite{Lafore98}
with a standard configuration using 3 imbricated domains with the innermost horizontal resolution equal to $\Delta$X=0.5km.
The results are already very encouraging, especially in the free atmosphere, with a good prediction of the meteorological 
parameters important to the optical turbulence estimation (temperature, wind speed).
The main discrepancies concerning the meteorological parameters between Meso-NH predictions and observations 
have been identified near the surface (between 2~m and 30~m above the ground), for the wind speed.
To overcome these specific limitations of the standard configuration, we discuss in this study the impact of the use of 
a higher horizontal resolution numerical configuration. 
Two domains with a horizontal resolution equal for both to $\Delta$X=0.1km, centered above Cerro Paranal and Cerro Armazones, have been 
added.
It is very important to well predict meteorological parameters such as wind speed and temperature, and their temporal evolution. 
Indeed, the intensity of the optical turbulence, characterized by the structure constant of the refractive index $C_N^2$,          
mainly depends on their values and gradients.
Our team have already used so far the Meso-NH model to perform some meteorological, 
and optical turbulence, numerical simulations, at other sites ( 
San Pedro Martir \cite{mas04}, Mount Graham \cite{hag10,hag11}, Antarctica \cite{las09,las10,las11}) discussing 
its abilities in predicting meteorological and astroclimatic parameters, but with less rich data samples.

\section{OBSERVATIONS DATA-SET} 
Two sites are investigated in this study: Cerro Paranal, site of the ESO Very Large Telescope (VLT), 
and Cerro Armazones, future site of the ESO European Extremely Large Telescope (E-ELT).
At Paranal, observations of meteorological parameters near the surface come from an automated weather station (AWS) and a 30~m high 
mast including a number of sensors at different heights. 
Both instruments are part of the VLT Astronomical Site Monitor \cite{vlt99}. 
Absolute temperature data are available at 2~m and 30~m above the ground.
Wind speed data are available at 10~m and 30~m above the ground. 
At Armazones, observations of the meteorological parameters near the ground surface come from the 
Site Testing Database \cite{Schock09}, more precisely from an AWS and a 30~m Tower (with temperature sensors 
and sonic anemometers). Data on temperature and wind speed are available at 2~m, 11~m, 20~m and 28~m above the ground.
At 2~m, at Armazones, temperature measurements from the AWS and the sonic anemometers are both available but we consider only those 
from the tower (accuracy of 0.1$^{o}C$ \cite{Skid07}). 
Those from the AWS are not reliable because of some drift effects (T. Travouillon, private communication). 
Wind speed observations are taken from the AWS (at 2 m) and from the sonic anemometers of the tower (at 11~m, 20~m and 28~m).
The outputs are sampled at 1 minute intervals.

\begin{figure}
\begin{center}
\begin{tabular}{c}
\includegraphics[width=0.80\textwidth]{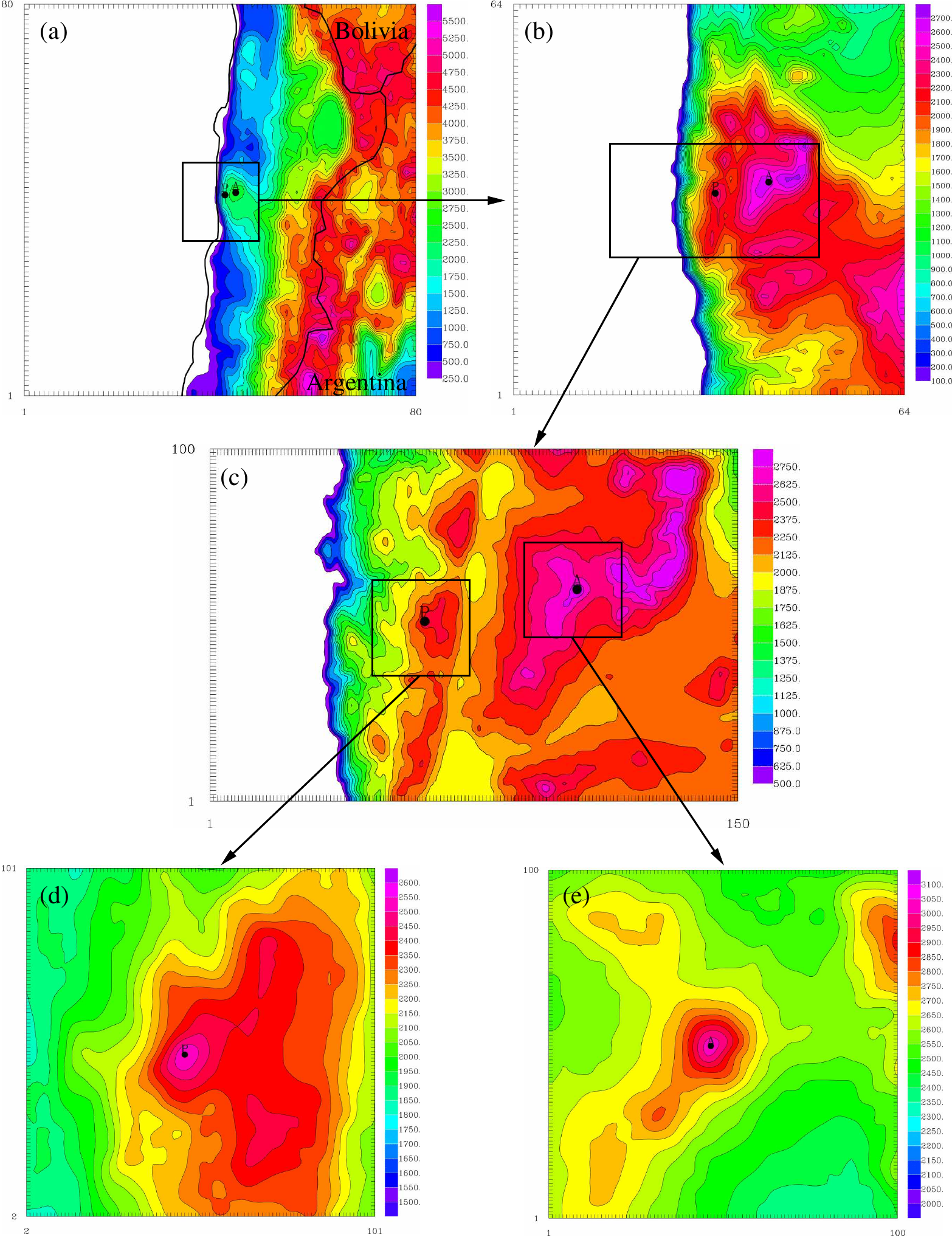}
\end{tabular}
\end{center}
\caption[oro]{\label{fig:oro} Orography (altitude in m) of the region of interest 
as seen by the Meso-NH model (polar stereographic projection)
for all the imbricated domains of the grid-nested configuration.
{\bf (a)} Domain 1 (orographic data from GTOPO),
{\bf (b)} Domain 2 (orographic data from GTOPO),
{\bf (c)} Domain 3 (orographic data from ISTAR),
{\bf (d)} Domain 4 (orographic data from ISTAR),
{\bf (e)} Domain 5 (orographic data from ISTAR),
A dot stands for Cerro Armazones and P dot stands for Cerro Paranal.
See Table \ref{tab:config} for the specifications of the domains (number of grid-points, domain extension, horizontal resolution).}
\end{figure}

\section{MESO-NH MODEL CONFIGURATION}
All the numerical simulations of the nights presented in this study are performed with the mesoscale numerical weather model 
Meso-NH \cite{Lafore98}.
The Meso-Nh model can simulate the temporal evolution of three-dimensional meteorological 
parameters over a selected finite area of the globe.
The system of hydrodynamic equations is based upon an anelastic formulation allowing for an effective filtering of acoustic waves.
It uses the Gal-Chen and Sommerville\cite{Gal75} coordinates system on the vertical and the C-grid in the formulation of
Arakawa and Messinger\cite{Arakawa76} for the spatial digitalization.
It employs an explicit three-time-level leap-frog temporal scheme with a time filter \cite{Asselin72}.
It employs a one-dimensional 1.5 turbulence closure scheme \cite{Cuxart00}. 
For this study we use a 1D mixing length proposed by Bougeault and Lacarr\`ere \cite{Bougeault89}.
The surface exchanges are computed using ISBA \cite{Noilhan89} (Interaction Soil Biosphere Atmosphere).
The grid-nesting technique \cite{Stein00}, employed in our study, consists of using different imbricated domains 
of the Digital Elevation Models (DEM i.e orography) extended on smaller and smaller surfaces, with increasing horizontal
resolution but with the same vertical grid. 
Two different grid-nesting configurations have been employed.
The standard configuration includes three domains (domains 1, 2 and 3 from Figure \ref{fig:oro}         
and Table \ref{tab:config}) with the lowest horizontal resolution equal to 10~km and the highest horizontal resolution equal to 0.5~km.
The second configuration is made of five imbricated domains, the first same three as the previous configuration, and other two
centered at both Paranal and Armazones sites, with a horizontal resolution of $\Delta$X=0.1 km (thereafter "5dom" configuration -
domains 1, 2, 3, 4 and 5 from Figure \ref{fig:oro}).
One can notice that using these configurations, we are able to do the forecast at both sites simultaneously.
The orographic DEMs we used for this project are the GTOPO\footnote{$http://www1.gsi.go.jp/geowww/globalmap-gsi/gtopo30/gtopo30.html$} 
with an intrinsic horizontal resolution of 1~km (used for the domains 1 and 2) and the ISTAR\footnote{Bought by ESO at the ISTAR Company 
- Nice-Sophia Antipolis, France} with an intrinsic horizontal resolution of 0.5~km (used for the domain 3, 4 and 5). 
Along the z-axis we have 62 levels distributed as follows: a first vertical grid point equal to 5~m, 
a logarithmic stretching of 20~$\%$ up to 3.5~km above the ground, and an almost constant vertical grid size of $\sim$600~m up to 23.8~km. 
The model has been parallelized using OPEN-MPI-1.4.3 and it run on local workstations as well as on the HPCF cluster of the European 
Centre for Medium weather Forecasts (ECMWF). 
The second solution permitted us to achieve relatively rich statistical estimates of these analysis.
\par
All simulations were initialized the day before at 18~UT and forced every 6 hours with the analyses from the ECMWF, 
and finished at 09~UT of the simulated day (duration: 15 hours).
The statistics we deal with in the next sections was computed only during night time, from 00~UT to 09~UT.
 
\begin{table}[t]
\begin{center}
\caption[config]{\label{tab:config} Meso-NH model configuration. 
In the second column the  horizontal resolution $\Delta$X, in the third column the number of 
grid points and in the fourth column the horizontal surface covered by the model domain.}
\begin{tabular}{|c|c|c|c|}
\hline
Domain & $\Delta$X & Grid Points & Surface \\
       & (km)      &             & (km$\times$km) \\
\hline
Domain 1 & 10  &  80$\times$ 80& 800$\times$800\\
Domain 2 & 2.5 &  64$\times$ 64& 160$\times$160\\
Domain 3 & 0.5 & 150$\times$100&  75$\times$50\\
Domain 4 & 0.1 & 100$\times$100&  10$\times$10\\
Domain 5 & 0.1 & 100$\times$100&  10$\times$10\\
\hline
\end{tabular}
\end{center}
\end{table}

\begin{table}[t]
\begin{center}
\caption[deltah]{\label{tab:deltah} Peak resolution with respect to the orographic model and the Meso-NH horizontal resolution.
GTOPO is the standard digital elevation model (DEM), from the US Geological Survey, used by the Meso-NH model.
It is not used in this study for the high-resolution domains (from domain 3 to 5), but we remind it here for comparisons purpose.
ISTAR is the DEM used in this study for both 3dom and 5dom configurations in the innermost domains.
SRTM (Shuttle Radar Topography Mission) is a more recent DEM with a higher intrinsic resolution \cite{srtm07}, 
that we planned to implement in further studies.
$\Delta$h represents the difference between the real altitude of the peak and the altitude of the peak in Meso-NH.}
\begin{tabular}{|c|c|c|c|c|}
\hline
Orographic & Intrinsic  & Meso-NH               & Paranal   & Armazones \\
model      & resolution & horizontal resolution & $\Delta$h & $\Delta$h \\
\hline
GTOPO      & 1km        & 500m                  & 204m      & 245m      \\
ISTAR      & 500m       & 500m (3dom)           & 156m      & 164m      \\
ISTAR      & 500m       & 100m (5dom)           & 89m       & 55m       \\
SRTM       & 90m        & 100m                  & 5m        & 26m       \\
\hline 
\end{tabular}
\end{center}
\end{table}

\section{COMPARISON MESO-NH OUPTUT vs OBSERVATIONS}
In a joint paper \cite{mas12} our team presented the results of Meso-NH simulations in the standard configuration for simulations 
at Cerro Paranal and Cerro Armazones.
The results were already very encouraging. 
However, the performances concerning the prediction of the meteorological parameters in proximity of the ground were mitigated. 
The statistics was performed on 20 nights, using two statistical operators beside the average: the bias and the root mean square 
error (RMSE). 
The prediction of the absolute temperature was very good.
At the 2~m and 30~m levels for Cerro Paranal, and at the 2~m, 11~m, 20~m and 28~m levels at Cerro Armazones, the biases between 
observations and model output were inferior to 1$^{o}C$.
The RMSE was equal, or slightly inferior to, $\sim$1$^{o}C$.
These are very good values.
On the contrary, the prediction of the wind speed above the surface was strongly underestimated by the model with respect to the 
observations, especially at the closest levels near the ground.
For example, at Cerro Armazones the bias reached $\sim$-3.6~$m{\cdot}s^{-1}$ at 2~m, with a RMSE equal to $\sim$4.6~$m{\cdot}s^{-1}$.
\par
To quantify the gain in statistic reliability, obtained with a very high horizontal resolution, 
in reconstructing surface meteorological parameters, the bias and the RMSE 
of the nightly temporal evolutions of temperature and wind speed have been computed, at each level where observations were available.
In the study presented here, a set of 9 nights (for Cerro Paranal) and 5 nights (for Cerro Armazones) 
have been simulated with the Meso-NH model both with the 3dom and 5dom configurations presented in the previous section.
This is a preliminary and ongoing work on the influence of the very high horizontal resolution ($\Delta$X=0.1km) on the prediction 
of meteorological parameters in proximity of the ground. 
That's why the number of investigated nights is not the same than in Masciadri \& Lascaux \cite{mas12}, 
and different for the two sites (it also depends on the available observations for a given day). 
However it is planned to extend it to better validate the results. 
\subsection{Wind and temperature temporal evolutions - mean, bias and RMSE}
\begin{figure}
\begin{center}
\begin{tabular}{c}
\includegraphics[width=0.95\textwidth]{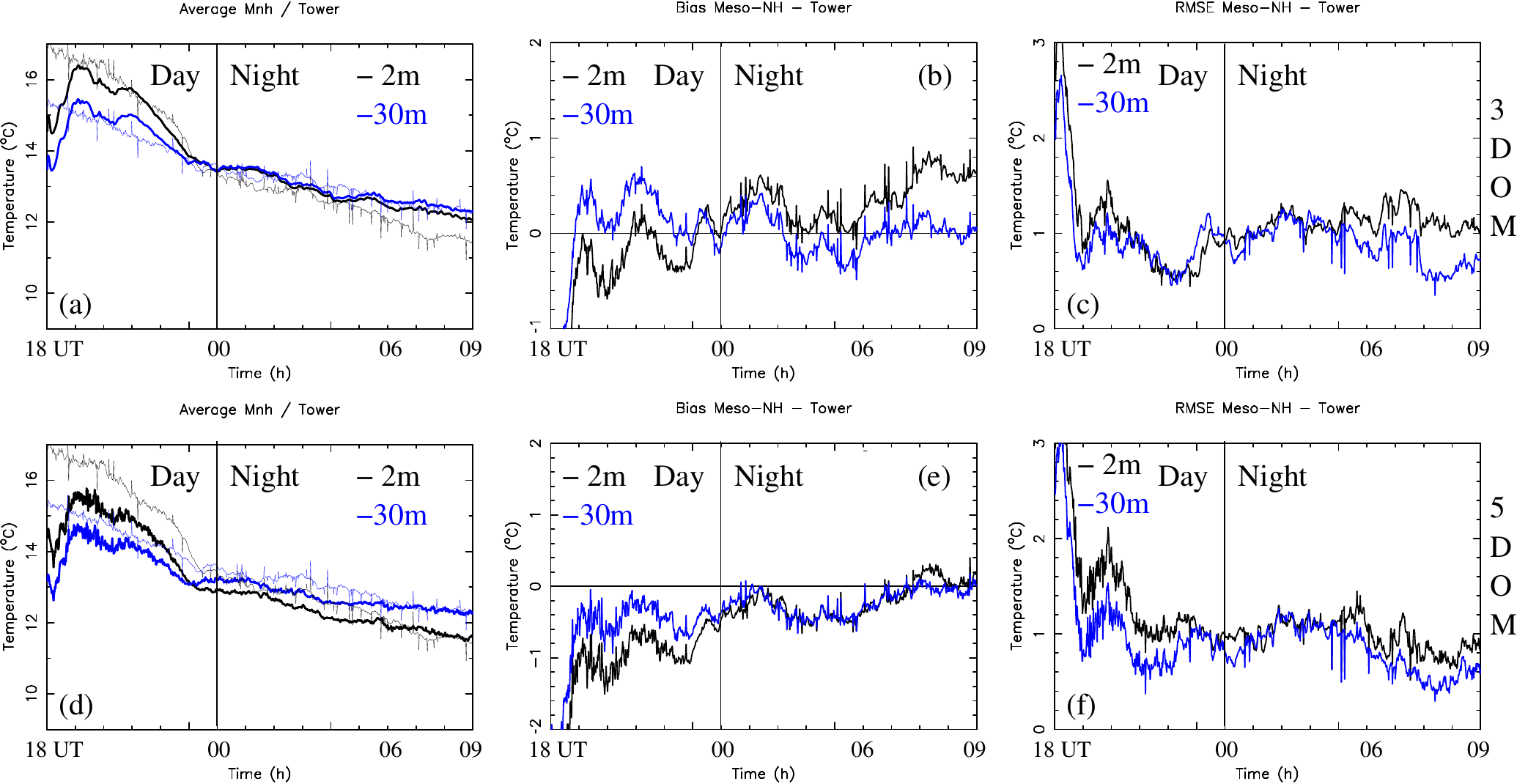}\\
\end{tabular}
\end{center}
\caption[tev_temp_par]{\label{fig:tev_temp_par} 9 nights statistics on the temporal evolution of the {\bf temperature} (in $^oC$) 
at {\bf Cerro Paranal} at 2~m (in black) and at 30~m (in blue).
{\bf (a)} average with the 3dom configuration,
{\bf (b)} bias with the 3dom configuration,
{\bf (c)} rmse with the 3dom configuration,
{\bf (d)} average with the 5dom configuration,
{\bf (e)} bias with the 5dom configuration,
{\bf (f)} rmse with the 5dom configuration.
The thin lines are observations averages. 
The thick lines are Meso-NH averages.}
\end{figure}

\begin{figure}
\begin{center}
\begin{tabular}{c}
\includegraphics[width=0.95\textwidth]{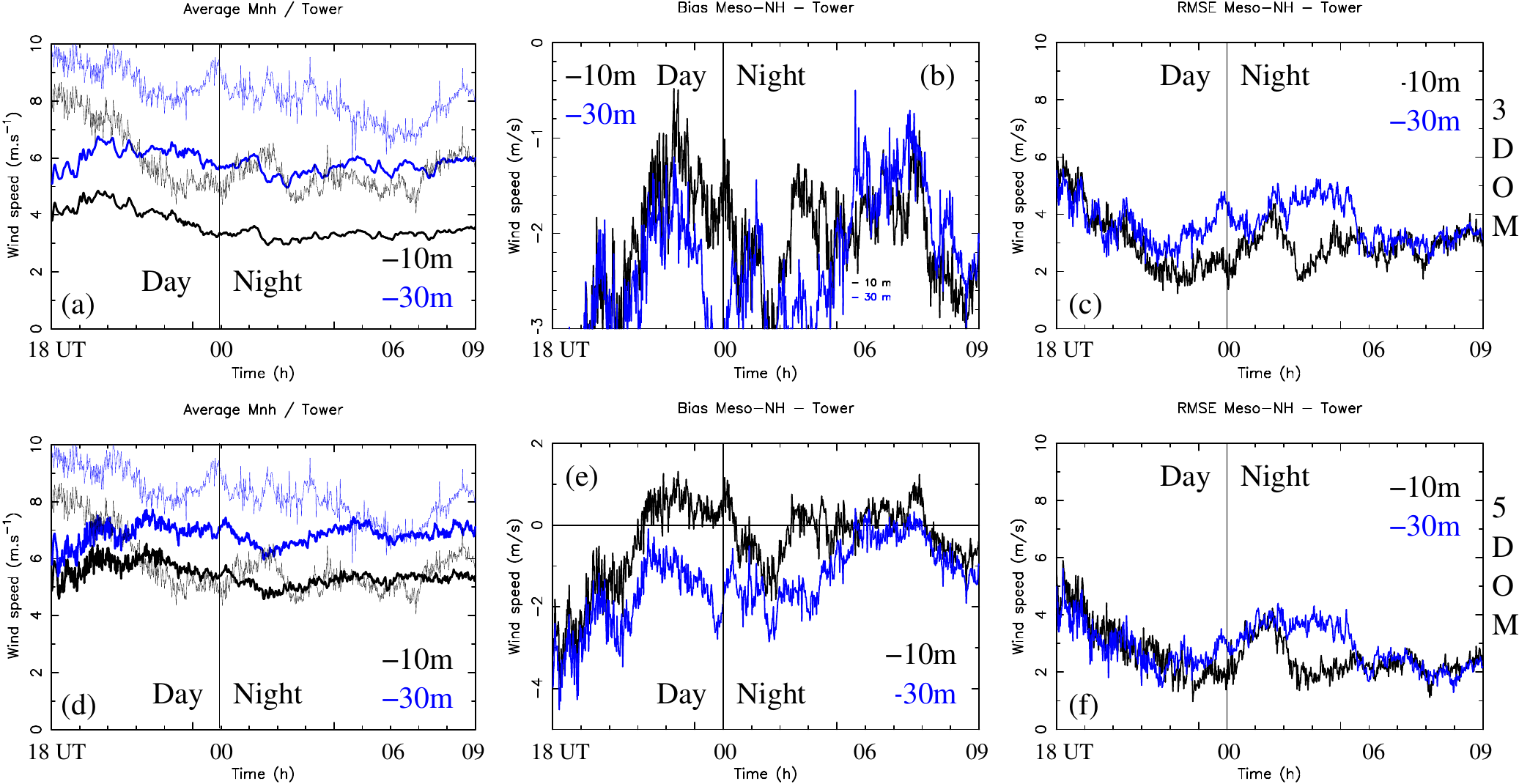}\\
\end{tabular}
\end{center}
\caption[tev_wind_par]{\label{fig:tev_wind_par} Same as Figure \ref{fig:tev_temp_par}, but for {\bf wind speed} (unit in $m{\cdot}s^{-1}$)
at {\bf Cerro Paranal}.}
\end{figure}

\begin{figure}[h]
\begin{center}
\begin{tabular}{c}
\includegraphics[width=0.95\textwidth]{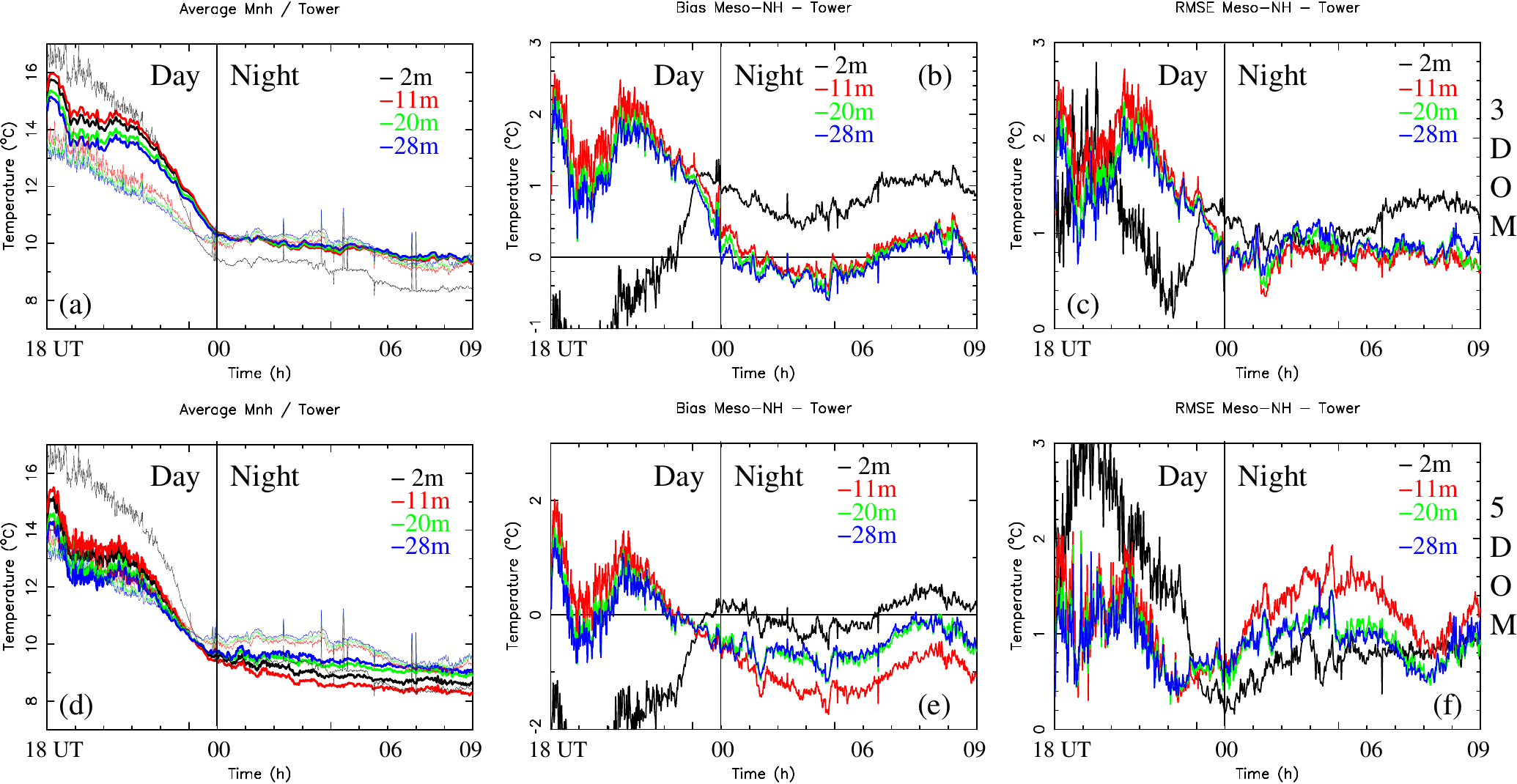}\\
\end{tabular}
\end{center}
\caption[tev_temp_arm]{\label{fig:tev_temp_arm} Same as Figure \ref{fig:tev_temp_par}, but for {\bf temperature} (in $^oC$) at 
{\bf Cerro Armazones}.}
\end{figure}

\begin{figure}[h]
\begin{center}
\begin{tabular}{c}
\includegraphics[width=0.95\textwidth]{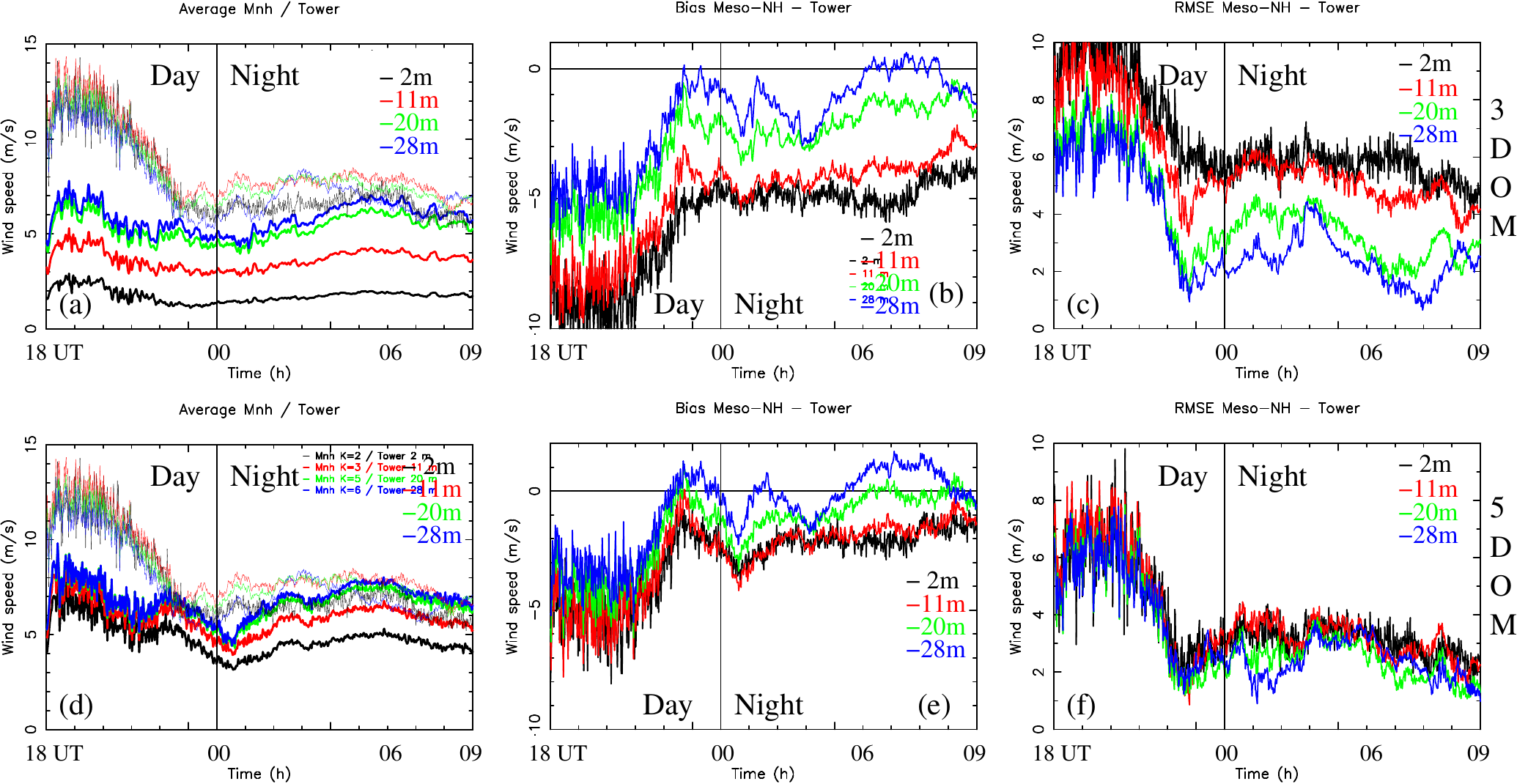}\\
\end{tabular}
\end{center}
\caption[tev_wind_arm]{\label{fig:tev_wind_arm} Same as Figure \ref{fig:tev_temp_arm}, but for {\bf wind speed} (unit in $m{\cdot}s^{-1}$) 
at {\bf Cerro Armazones}.}
\end{figure}

In this section we present the average, bias and RMSE of the nightly temporal evolution of wind speed and temperature at 
Cerro Paranal and Cerro Armazones.
The average, the bias and the RMSE between simulated and observed parameters are defined as:\\
$$
AVERAGE = \sum_{i=1}^{N}\frac{X_{i}}{N}
$$
$$
BIAS = \sum_{i=1}^{N} \frac{(Y_{i}-X_{i})}{N} 
$$
$$
RMSE = \sum_{i=1}^{N} \sqrt{\frac{(Y_{i}-X_{i})^{2}}{N}}
$$
with $X_{i}$ beeing the observed values and $Y_{i}$ the simulated values.
$N$ is the total number of times in which the couple ($X_{i}$,$Y_{i}$) is available and different from zero.
In the case of the comparisons of the temporal evolutions (Figures \ref{fig:tev_temp_par}, \ref{fig:tev_wind_par},
\ref{fig:tev_temp_arm} and \ref{fig:tev_wind_arm}),
$N \leq n$, with $n$ the number of nights, which simply means that observed values are not always 
available at every minutes for every nights and in these cases the statistics is performed on a number of points inferior to the total 
number of nights.
In the case of Tables \ref{tab:bias_rmse_temp} and \ref{tab:bias_rmse_wind}, N represents the total number of available measurements, for 
all the nights considered.\\
The temporal evolutions of the average, bias and RMSE for both 3dom and 5dom configurations, are displayed on 
Figures \ref{fig:tev_temp_par} and \ref{fig:tev_wind_par} at Cerro Paranal (temperature and wind speed, respectively) and 
Figures \ref{fig:tev_temp_arm} and \ref{fig:tev_wind_arm} at Cerro Armazones (temperature and wind speed, respectively).\\
A summary of the values of bias and RMSE are reported on Tables \ref{tab:bias_rmse_temp} and \ref{tab:bias_rmse_wind}, for temperature and
wind speed, respectively.
It permits us to quantify the gain obtained in the prediction of the temperature and wind speed near the ground, going from an
horizontal resolution equal to $\Delta$X=0.5km (3dom) to $\Delta$X=0.1km (5dom).
We focus our attention on nightly values only, as in Masciadri \& Lascaux \cite{mas12}.
The first 6 hours of simulation, during daytime, are affected by sone spurious values due to the adaption of the model to the orography, 
especially during the first part of the simulation.
\par
The first important thing to notice is that the statistics on temperature remains good with the 5dom configuration, as it was for 
the 3dom configuration.
We have biases inferior, or very close to, 1$^{o}C$, and RMSEs close to 1$^{o}C$.
Concerning the temperature prediction by the model, one important feature was better reproduced with the 5dom configuration than with 
the 3dom configuration.
The observed temperature gradient during night time between 2~m and the successive levels was almost absent with the 3dom configuration
(Figures \ref{fig:tev_temp_par}a and \ref{fig:tev_temp_arm}a).
The 5dom configuration was able to better reproduce it, at Cerro Paranal and at Cerro Armazones 
(Figures \ref{fig:tev_temp_par}d and \ref{fig:tev_temp_arm}d).
In particular the prediction of the inversion of the vertical gradient of temperature near the surface, during the passage from 
day to night (from negative to positive), is improved in the 5dom simulations.
However, as it was shown for the standard  3dom configuration \cite{mas12}, the 11~m level at Armazones still presents a wrong 
gradient tendency with the 5dom configuration. 
\par
Concerning the wind speed, there was a strong understimation of its intensity with the standard 3dom configuration,
especially at the first 2~m level.
It seems that the Meso-NH model isn't able to reproduce high wind in proximity of the ground, at the peaks, with the 3dom configuration, 
leading to high values of bias and RMSE.
This problem is partially solved with the 5dom configuration, where the model is now able to produce high winds even for the first 
levels of the model at Cerro Paranal and Cerro Armazones.
At 10~m at Cerro Paranal for example, the bias is strongly reduced (cf. Table \ref{tab:bias_rmse_wind}), from -2.07~$m{\cdot}s^{-1}$ to 
-0.16~$m{\cdot}s^{-1}$.
Biases at 30~m at Cerro Paranal, at 2~m, 11~m and 20~m at Cerro Armazones are more than halved, having now values smaller than for the 3dom 
configuration: for the 3dom configuration, they are between -2~$m{\cdot}s^{-1}$ and -5~$m{\cdot}s^{-1}$; for the 5dom configuration 
they are between -0.8~$m{\cdot}s^{-1}$ and -2~$m{\cdot}s^{-1}$.
At 28~m at Cerro Armazones, the bias is even equal to 0~$m{\cdot}s^{-1}$, whereas it was close to -1~$m{\cdot}s^{-1}$ with the 3dom 
configuration.
The RMSE values are reduced too. The strongest RMSE, at 2~m at Cerro Armazones, was equal to 5.72~$m{\cdot}s^{-1}$ with the 3dom 
configuration, it is now equal to 3.15~$m{\cdot}s^{-1}$ with the 5dom configuration. 
However, even though the prediction of the wind speed with the 5dom configuration is very good, 
one minor discrepancy still remains at Cerro Armazones. 
In the observations, a low level jet is visible at 11~m. This level exhibits the 
highest wind speed with respect to the others levels at 2~m, 20~m and 28~m.
On the contrary, the wind speed is constantly increasing in the model from 2~m to 28~m, suggesting that the low level jet may be 
at slightly higher levels in the model than in the observations.   
\begin{table}
\begin{center}
\caption[bias_rmse_temp]{\label{tab:bias_rmse_temp} Bias and RMSE of the {\bf temperature} between
Meso-NH simulations (3dom and 5dom configurations) and observations, at Cerro Paranal (left) and Cerro Armazones (right).
Units in $^oC$. Computations are made with 9 nights for Paranal and 5 nights for Armazones. }
\begin{tabular}{|c|c|c|c||c|c|c|c|c|c|}
\hline
\multicolumn{2}{|c|}{PARANAL}& \multirow{2}{*}{2~m}& \multirow{2}{*}{30 m}& \multicolumn{2}{|c|}{ARMAZONES} &\multirow{2}{*}{2~m}&\multirow{2}{*}{11m}&\multirow{2}{*}{20~m}&\multirow{2}{*}{28~m} \\
\multicolumn{2}{|c|}{(9 nights)}& & &\multicolumn{2}{|c|}{(5 nights)} & & & &\\
\hline
\multirow{2}{*}{3dom} & BIAS     & 0.37  & 0.00   & \multirow{2}{*}{3dom} & BIAS              & 0.84  & 0.06   & -0.03 & -0.08 \\
                      & RMSE     & 1.13  & 0.92   &                       & RMSE              & 1.11  & 0.75   & 0.83  & 0.87  \\
\hline
\multirow{2}{*}{5dom} & BIAS     & -0.20 & -0.22  & \multirow{2}{*}{5dom} & BIAS              & 0.02  & -1.04  & -0.56 & -0.53 \\
                      & RMSE     & 1.00  & 0.85   &                       & RMSE              & 0.76  & 1.30   & 0.95  & 0.94  \\
\hline
\end{tabular}
\end{center}
\end{table}
\begin{table}
\begin{center}
\caption[bias_rmse_wind]{\label{tab:bias_rmse_wind} Same as Table \ref{tab:bias_rmse_temp} but for the {\bf wind speed}.
Units in $m{\cdot}s^{-1}$.}
\begin{tabular}{|c|c|c|c||c|c|c|c|c|c|}
\hline
\multicolumn{2}{|c|}{PARANAL}& \multirow{2}{*}{10~m}& \multirow{2}{*}{30 m}& \multicolumn{2}{|c|}{ARMAZONES} &\multirow{2}{*}{2~m}&\multirow{2}{*}{11m}&\multirow{2}{*}{20~m}&\multirow{2}{*}{28~m} \\
\multicolumn{2}{|c|}{(9 nights)}& & &\multicolumn{2}{|c|}{(5 nights)} & & & &\\
\hline
\multirow{2}{*}{3dom} & BIAS     & -2.07 & -2.16  & \multirow{2}{*}{3dom} & BIAS              & -4.78  & -3.95  & -1.97 & -0.92 \\
                      & RMSE     & 2.94  & 3.77   &                       & RMSE              & 5.72   & 5.10   & 3.34  & 2.49  \\
\hline
\multirow{2}{*}{5dom} & BIAS     & -0.16 & -1.01  & \multirow{2}{*}{5dom} & BIAS              & -2.06 & -1.92  & -0.78 & 0.01  \\
                      & RMSE     & 2.46  & 2.97   &                       & RMSE              & 3.15  & 3.21   & 2.50  & 2.49  \\
\hline
\end{tabular}
\end{center}
\end{table}
\clearpage
\clearpage
\subsection{Wind and temperature temporal evolutions - 11 November 2007}
\begin{figure}[t]
\begin{center}
\begin{tabular}{c}
\includegraphics[width=0.65\textwidth]{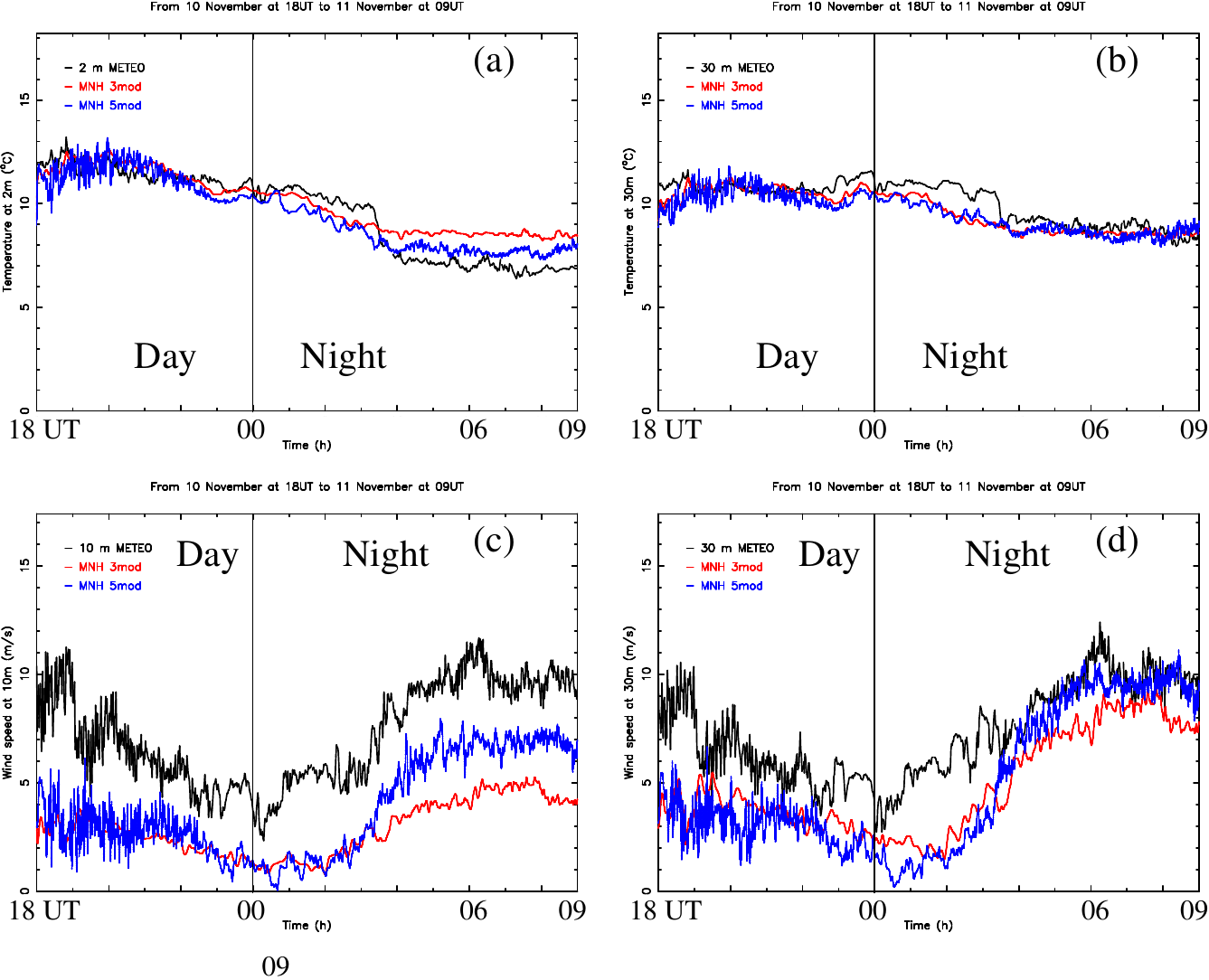}
\end{tabular}
\end{center}
\caption[par_111107_t_w]{\label{fig:par_111107_t_w} Temporal evolutions of 
{\bf (a)} {\bf temperature} at 2~m, 
{\bf (b)} {\bf temperature} at 30 m, 
{\bf (c)} {\bf wind speed} at 10~m and 
{\bf (d)} {\bf wind speed} at 30 m, at {\bf Cerro Paranal}, for the night 11 November 2007. 
Units in $^oC$ for the temperature and in $m{\cdot}s^{-1}$ for the wind speed.
In black, observations. In red, the 3dom simulation. In blue, the 5dom simulation.}
\end{figure}

\begin{figure}
\begin{center}
\begin{tabular}{c}
\includegraphics[width=0.65\textwidth]{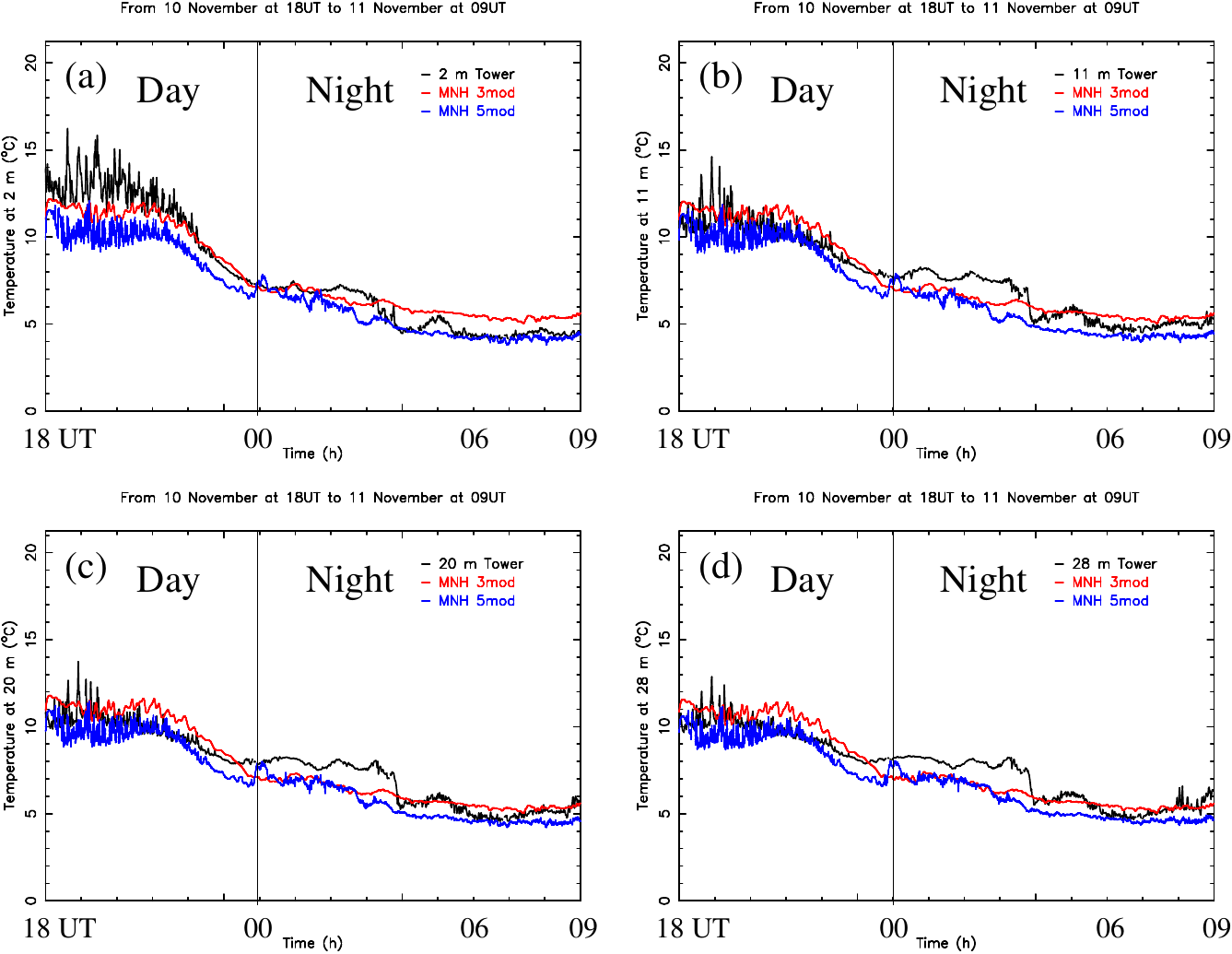}
\end{tabular}
\end{center}
\caption[arm_111107_t]{\label{fig:arm_111107_t} Temporal evolutions of {\bf temperature} 
{\bf (a)} at 2~m, 
{\bf (b)} at 11~m, 
{\bf (c)} at 20~m and 
{\bf (d)} at 28~m, at {\bf Cerro Armazones}, for the night 11 November 2007. 
Units in $^oC$. 
In black, observations. In red, the 3dom simulation. In blue, the 5dom simulation.}
\end{figure}

\begin{figure}[h]
\begin{center}
\begin{tabular}{c}
\includegraphics[width=0.65\textwidth]{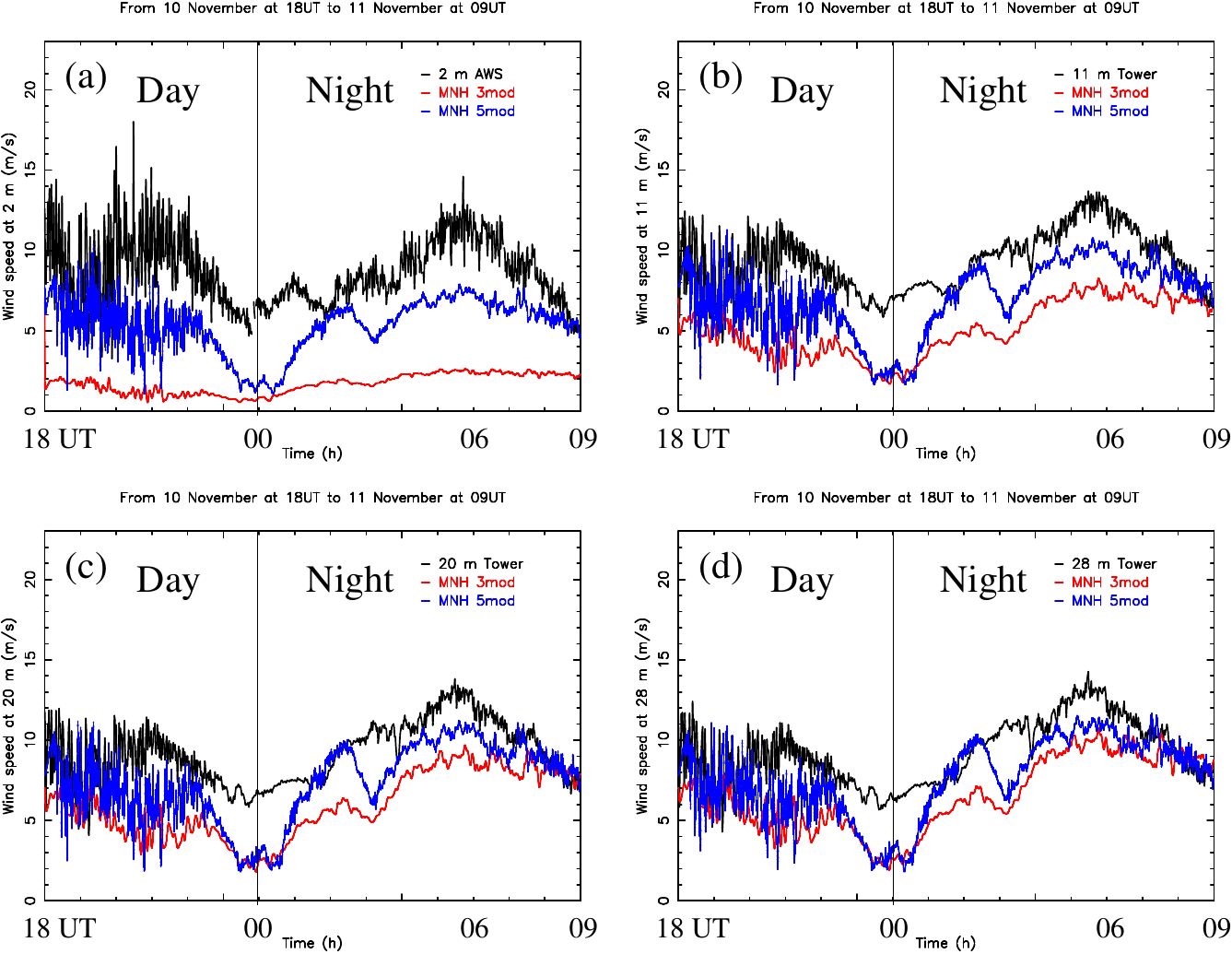}
\end{tabular}
\end{center}
\caption[arm_111107_w]{\label{fig:arm_111107_w} Temporal evolutions of {\bf wind speed} 
{\bf (a)} at 2~m, 
{\bf (b)} at 11~m, 
{\bf (c)} at 20~m and 
{\bf (d)} at 28~m, at {\bf Cerro Armazones}, for the night 11 November 2007.                       
Units in $m{\cdot}s^{-1}$.
In black, observations. In red, the 3dom simulation. In blue, the 5dom simulation.}
\end{figure}

In this section we present the results for one night only, the 11 November 2007.
It allows us to illustrate how good can be the improvements on the prediction of wind speed and temperature near the ground,
for one single night, if we use a higher horizontal resolution.
Figure \ref{fig:par_111107_t_w} displays the temporal evolution of the temperature (top) and the wind speed (bottom) for this night, at 
Cerro Paranal, for both 3dom and 5dom configurations, and observations, at 2~m and 30 m for the temperature and at 10~m and 30 m 
for the wind speed.
Figure \ref{fig:arm_111107_t} displays the temporal evolution of the temperature at Cerro Armazones for the same night, at four 
different levels (2~m, 11~m, 20~m, 28~m).
Figure \ref{fig:arm_111107_w} displays the temporal evolution of the wind speed at Cerro Armazones for the same night, at four
different levels (2~m, 11~m, 20~m, 28~m).
At 2~m, the forecasted nightly temperature with the 3dom configuration was slightly warmer than the observations, for both sites.
The difference is very small, smaller than 1.5$^oC$ at Paranal and 1$^oC$ at Armazones.
At 30~m, at the end of the night, there is almost no bias between observations and forecast.
Using the 5dom configuration, a small cooling appears with respect to the 3dom configuration, at 2~m, making the prediction almost perfect 
at this level.
More over, the prediction at other levels m is not degraded 
(except for a very small cooling with respect to the observations at 11~m, 20 and 28~m at Cerro Armazones).
\par
The major improvements brought by the highest horizontal resolution configuration can be seen for the wind speed.
At Paranal, for this night, the wind speed reached around 10~$m{\cdot}s^{-1}$ at 2~m and 10~m.
With the 3dom configuration, the wind speed at 2~m stalled at 5~$m{\cdot}s^{-1}$, whereas with the 5dom configuration it reached 
7~$m{\cdot}s^{-1}$, closer to the observed 10~$m{\cdot}s^{-1}$.
The prediction at 30 m with the 3dom configuration was satisfactory (around 8~$m{\cdot}s^{-1}$), but it is almost perfect at the end 
of the night with the 5dom configuration where it reached 10~$m{\cdot}s^{-1}$.
The improvement is even more significative at Cerro Armazones at the first observed level (2~m). 
The observed wind speed oscillated during the night between 5 and 12~$m{\cdot}s^{-1}$.
With the 3dom configuration, the first simulated level forecasted a wind speed always inferior at 2~$m{\cdot}s^{-1}$. 
Using the 5dom configuration, the forecasted wind oscillated between 5 and 8~$m{\cdot}s^{-1}$ (2~m).
More over, the temporal evolution exhibits a peak of intensity at the exact same time (around 05:30 UT).
This improvement in the wind speed forecast is visible also at the others 3 levels (11~m, 20~m, 28~m).
\clearpage
\section{CONCLUSION}
The intensity of the optical turbulence is mainly driven by the gradients of some meteorological parameters 
(wind speed, temperature) that determine the vertical profiles of $C_N^2$.
As these vertical profiles of $C_N^2$, from which derive all the
astroclimatic parameters useful for astronomers (seeing, wavefront coherence time, isoplanatic angle...)
are characterized by a maximum in the surface layer, a good knowledge and prediction of the surface meteorological parameters
is mandatory to obtain good forecast of the optical turbulence at a given site.
In this study we have investigated the impact of very high horizontal resolution on the prediction of meteorological parameters 
(temperature and wind speed) near the surface at two montainous sites: Cerro Paranal (site of the VLT) and Cerro Armazones 
(future site of the E-ELT).
In the standard configuration \cite{mas12}, less demanding in computing resources, the results are already very encouraging over the entire 
atmosphere above the sites of interest. 
Nevertheless, some discrepancies on the wind speed prediction near the ground were present, with low-level simulated winds lower 
than the obervations.
The work presented in this study adressed this particular problem by testing a very high horizontal resolution configuration 
(5dom, with $\Delta$X=0.1km for the innermost domains), more demanding in computing resources.
In this preliminary study (only 9 nights for Cerro Paranal and 5 nights for Cerro Armazones were investigated), 
we demonstrated that a very high horizontal resolution improved significatively the performance of the 
model in proximity of the ground.
Concerning the temperature, not only the prediction of the low-level temperature 
remained very good (with biases inferior to 1$^{o}C$ at all levels at both sites, 
and even equal to 0$^{o}C$ at 2 m at Cerro Paranal), but the day to night temperature gradient inversion was more accurately reproduced.
Concerning the wind speed, the comparison between 5dom simulations and observations gave biases reduced by more than a half at all levels, 
at both sites, with respect to the standard 3dom configuration.
\\

\clearpage 
\acknowledgments     
Meteorological data-set from the Automatic Weather Station (AWS) and mast at Cerro Armazones are from the 
Thirty Meter Telescope Site Testing - Public Database Server\cite{Schock09}.
Meteorological data-set from the AWS and mast at Cerro Paranal are from ESO Astronomical Site Monitor 
(ASM - Doc.N. VLT-MAN-ESO-17440-1773). 
We are very greatful to the whole staff of the TMT Site Testing Working Group for providing information about 
their data-set as well as to Marc Sarazin for his constant support to this study and for providing us the ESO data-set used in this study. 
Simulations are run partially on the HPCF cluster of the European Centre for Medium Weather Forecasts (ECMWF) - Project SPITFOT. 
This study is co-funded by the ESO contract: E-SOW-ESO-245-0933.
 


\begin{thebibliography}{1}

\bibitem{mas12}
E. Masciadri and F. Lascaux, "MOSE: a feasibility study for optical turbulence forecast with the MesoNh mesoscale model
to support AO facilities at ESO sites (Paranal, Armazones)", {\em SPIE Astronomical Telescopes and Instrumentation}, Amsterdam, 
1-6 July, 2012.
\bibitem{Lafore98}
J. P. Lafore, J. Stein, N. Asencio, P. Bougeault, V. Ducrocq, J. Duron, C. Fischer, P. Hereil,
P. Mascart, V. Masson, J.-P. Pinty, J.-L. Redelsperger, E. Richard and J. Vil\`a-Guerau de Arellano,
"The Meso-Nh atmospheric simulation system. Part I: Adiabatic formulation and control simulations",
{\em Annales Geophysicae}, 16, pp.~90-109, 1998.
\bibitem{mas04}
E. Masciadri, R. Avila and L. J. S\'anchez, "Statistic Reliability of the Meso-Nh Atmospherical Model for 3D $C_N^2$ simulations", 
{\em Rev. Mex. Astron. Astrofisica}, 40, pp.~3-14, 2004.
\bibitem{hag10}
S. Hagelin, E. Masciadri and F. Lascaux, "Wind speed vertical distribution at Mt Graham", {\em MNRAS}, 407, pp.~2230-2240, 2010.
\bibitem{hag11}
S. Hagelin, E. Masciadri and F. Lascaux, "Optical turbulence simulations at Mt Graham using the Meso-NH model", {\em MNRAS}, 412, 
pp.~2695-2706, 2011.
\bibitem{las09}
F. Lascaux, E. Masciadri, S. Hagelin and J. Stoesz, "Mesoscale optical turbulence simulations at Dome C", {\em MNRAS}, 398, 
pp.~1093-1104, 2009.
\bibitem{las10}
F. Lascaux, E. Masciadri, S. Hagelin and J. Stoesz, "Mesoscale optical turbulence simulations at Dome C: refinements", {\em MNRAS}, 403, 
pp.~1714-1718, 2010.
\bibitem{las11}
F. Lascaux, E. Masciadri and S. Hagelin, "Mesoscale optical turbulence simulations above Dome C, Dome A and South Pole", {\em MNRAS}, 
411, pp.~693-704, 2011.
\bibitem{vlt99}
S. Sandrock, R. Amestica, "VLT Astronomical Site Monitor - ASM Data User Manual",
{\em Doc no.: VLT-MAN-ESO-17440-1773}, 1999.
\bibitem{Schock09}
M. Schoeck, S. Els, R. Riddle, W. Skidmore, T. Travouillon, R. Blum, E. Bustos, G. Chanan, S. G. Djorgovski, P. Gillett, 
B. Gregory, J. Nelson, A. Ot\'arola, J. Seguel, J. Vasquez, A. Walker, D. Walker and L. Wang, 
"Thirty Meter Telescope Site Testing I: Overview",
{\em PASP}, 121, pp.~384-395, 2009.
\bibitem{Skid07}
W. Skidmore, T. Travouillon and R. Riddle, "Report of the calibration of the T2-Armazones, 30-m tower air temperature 
sensors and sonic anemometers, the cross comparison of weather stations and sonic anemometers and turbulence measurements 
of sonic anemometers and finewire thermocouples", {\em Internal TMT Report}, 2007. 
\bibitem{Lipps82}
F. Lipps and R. S. Hemler, "A scale analysis of deep moist convection and some related numerical calculations",
{\em J. Atmos. Sci.}, 39, pp.~2192-2210, 1982.
\bibitem{Gal75}
T. Gal-Chen and C.J. Sommerville, "On the use of a coordinate transformation for the solution of the Navier-Stokes equations",
{\em J. Comput. Phys.}, 17, pp.~209-228, 1975.
\bibitem{Arakawa76}
A. Arakawa and F. Messinger, "Numerical methods used in atmospheric models",
{\em GARP Tech. Rep.}, 17, WMO/ICSU, Geneva, Switzerland, 1976.
\bibitem{Asselin72}
R. Asselin, "Frequency filter for time integration",
{\em Mon. Weather. Rev.}, 100, pp.~487-490, 1972.
\bibitem{Cuxart00}
J. Cuxart, P. Bougeault and J.-L. Redelsperger, "A turbulence scheme allowing for mesoscale and large-eddy simulations",
{\em Q. J. R. Meteorol. Soc.}, 126, pp.~1-30, 2000.
\bibitem{Bougeault89}
P. Bougeault and P. Lacarr\`ere, "Parameterization of orographic induced turbulence in a mesobeta scale model",
{\em Mon. Weather. Rev.}, 117, pp.~1972-1890, 1989.
\bibitem{Noilhan89}
J. Noilhan and S. Planton, "A simple paramterization of land surface processes for meteorological models",
{\em Mon. Weather. Rev.}, 117, pp.~536-549, 1989.
\bibitem{Stein00}
J. Stein, E. Richard, J.-P. Lafore, J.-P. Pinty., N. Asencio and S. Cosma, "High-Resolution Non-Hydrostatic Simulations of 
Flash-Flood Episodes with Grid-Nesting and Ice-Phase Parameterization",
{\em Meteorol. Atmos. Phys.}, 72, pp.~203-221, 2000.
\bibitem{srtm07}
G. Farr, et al., "The Shuttle Radar Topography Mission",
{\em Rev. Geophys.}, 45, RG2004, 2007.

\end{thebibliography}
\end{document}